\documentclass[conference]{IEEEtran}
\IEEEoverridecommandlockouts
\usepackage{cite}
\usepackage{amsmath,amssymb,amsfonts}
\usepackage{algorithmic}
\usepackage{algorithm2e}
\RestyleAlgo{ruled}
\usepackage{graphicx}
\usepackage{textcomp}
\usepackage{xcolor}
\usepackage{booktabs}
\usepackage{enumitem}
\def\BibTeX{{\rm B\kern-.05em{\sc i\kern-.025em b}\kern-.08em
    T\kern-.1667em\lower.7ex\hbox{E}\kern-.125emX}}

\usepackage{eqparbox,array}

\ifCLASSOPTIONcompsoc
  \usepackage[caption=false,font=normalsize,labelfont=sf,textfont=sf]{subfig}
\else
  \usepackage[caption=false,font=footnotesize]{subfig}
\fi

\usepackage{eqparbox,array}

\begin{document}

\title{Incomplete Multiview Learning via Wyner Common Information
}

\author{%
\IEEEauthorblockN{AbdAlRahman Odeh\IEEEauthorrefmark{1}, Teng-Hui Huang\IEEEauthorrefmark{2},  and Hesham El Gamal\IEEEauthorrefmark{3}}
\IEEEauthorblockA{\IEEEauthorrefmark{1}\IEEEauthorrefmark{2}%
School of Electrical and Computer Engineering, University of Sydney, NSW 2006, Australia}
\IEEEauthorblockA{%
\IEEEauthorrefmark{3}Faculty of Engineering, University of Sydney, NSW, Australia 2006
}
\IEEEauthorblockA{email:\{abdal-rahman.odeh,tenghui.huang,hesham.elgamal\}@sydney.edu.au}
}

\maketitle

\begin{abstract}
Incomplete multiview clustering is of high recent interest, fueled by the advancement of common information-based deep multiview learning. The practical scenarios where unpaired multiview data with missing values have wide applications in generative learning, cross-modal retrieval, and wireless device identification problems. Following the perspective that the shared information between the incomplete multiview data aligns with the cluster targets, recent works have generalized the well-known common information frameworks in information theory multiview learning problems, with improved performance reported. Different from previous works, we extend the frameworks to incomplete multiview clustering problems and propose an efficient solver: Wyner Incomplete MultiView Clustering (WyIMVC). Interestingly, the common randomness in WyIMVC allows for joint clustering and missing value inference in contrast to the compared methods in the literature. Moreover, leveraging the difference-of-convex structure of the formulated problems, we propose an efficient solver with a convergence guarantee independent of initialization. Empirically, our solver outperforms the state-of-the-art solvers in a range of incomplete multiview datasets with varying numbers of views and dimensions.
\end{abstract}

\begin{IEEEkeywords}
Wyner Common Information, Difference of Convex, Optimization, Algorithm
\end{IEEEkeywords}

\section{Introduction}
Multiview learning refers to a machine learning paradigm with the goal of enhancing the performance of data in multiple formats. A common belief in multiview learning is that it is beneficial to leverage the shared information between multiple views of observations for the task of interest, e.g., classification, clustering, information retrieval~\cite{chowdhury2010introduction,xu2015inmulti,multimodal2019survey,chen2020simple,do2021clustering,xu2022multi}. Following this perspective, when the task target coincides with the common randomness in multiview data, characterizing the shared information is important for solving the multiview learning problem.

Among recent works in common information-based multiview learning, the applications of well-known common information frameworks in information theory, e.g., Wyner~\cite{wyner1975common} and G\'acs-K\"ornor common information~\cite{gacs1973common}, to multiview learning tasks have been shown to be effective in formulating interpretable optimization objectives, scalable network architectures and rigorous theoretic foundations. In~\cite{huang2019universal} and~\cite{huang2020multimodal}, the connections between a local version of Wyner common information and Hirschfeld-Gebelein-Rényi maximal correlation are established. The proposed modal decomposition features algorithm demonstrated effectiveness for partially-supervised multiview learning settings. In approximating the G\'acs-K\"ornor common information,~\cite{salamatian2020approximate} relaxes the exchangeable Markov constraint to a ``straight-through'' one in bivariate settings, which allows for more general characterization of the correlation of bivariate sources as compared to the seminal framework. In the generalization of the definition of common randomness to continuous settings, the common information dimension is defined with closed-form expressions for the jointly Gaussian special case provided~\cite{diggavi2023cid}. Moreover, efficient common information extraction algorithms by adopting the theory of variational inference have demonstrated improved empirical performance in multiview representation learning~\cite{ryu2019learning,kleinman2023gacs,huang2024efficient}. However, most works in this line assume access to joint distributions of multiview observations, or a complete and paired dataset (no missing values for all views in each sample). Such an assumption might not apply to \textbf{practical incomplete multiview learning scenarios}, where there are unknown values for a subset of all views due to delay, corruption, or expensiveness in data collection~\cite{xu2015inmulti}.

Previous works in incomplete multiview clustering either adopt a two-step complete-and-cluster approach or a separate-and-align method. In the former case, the missing values are inferred from available views, followed by full view multiview clustering~\cite{lin2021completer,wen2021unified,lin2022dual,huang2024icasspcomplete}; whereas in the latter case, view-specific feature extraction or projection is performed, then paired samples are used to align these extracted features in the same low-dimensional subspaces, where clustering algorithms are applied to~\cite{deng2023projective,image2023app}. Different from these earlier two-step approaches, where clustering is performed after feature completion/extraction, we aim to address the challenge by directly characterizing the common randomness to fully leverage the shared information between available views. This is motivated by the insights in Wyner and G\'acs-K\"ornor common information~\cite{wyner1975common,wynerLossy11,gacs1973common,salamatian2020approximate}, i.e., statistical independence of view observations conditioned on common cluster targets, which in this context implies that inferring missing values and independence of view-specific features are by-products of clustering.

To our furthest knowledge, there are only a handful of previous works that adopt an information-theoretic formulation of common information-based incomplete multiview clustering. In contrastive learning based approaches, \cite{lin2021completer} focuses on the two-view case, but only uses complete view samples to maximize the correlation of view-specific features, while minimizing self-and-cross-view reconstruction losses to infer the missing values. While a follow-up work further generalizes the approach to more than two views~\cite{lin2022dual}, the methods simply concatenate all extracted features as a long vector for clustering. Therefore, they do not directly estimate the common randomness.~\cite{liu2024incomplete} proposed a three-step method that first forms view-specific similarity graphs, which are used for view-specific feature extraction, and jointly learns a common feature. The approach adopted the multiview information bottleneck method to maximize the relevance of the common feature~\cite{zaidi2020information,xu2014large}. However, it has limited scalability due to the step of forming the similarity matrices. Additionally, the common feature is Gaussian distributed and hence relies on an additional clustering algorithm for evaluation.

Our contributions are summarized as follows. First, we propose a novel information-theoretic formulation of incomplete multiview clustering extending the Wyner common information paradigm. Our formulation applies to arbitrary missing view indices settings without imputing the missing values. Second, we leverage the difference-of-convex (DC) structures in our formulations, which gives a closed-form characterization of the common randomness using the difference-of-convex algorithm (DCA). Owing to the insights, our solver directly predicts a cluster target without missing value imputation or the application of ad-hoc clustering algorithms as compared to previous methods. Lastly, we evaluate our solver on a range of benchmark incomplete multiview clustering datasets with varying numbers of views and feature dimensions. Our solver outperforms the state-of-the-art solvers in clustering accuracy over different percentages of missing values in the evaluated datasets.




\paragraph*{Notation}
The upper-case letter $Z$ denotes a random variable with density function $P_z$, whereas the lower-case letter $z$ refers to a realization $z\sim P_z$ with probability $p(z)$ and cardinality $|\mathcal{Z}|$. Boldface upper-case letters $\boldsymbol{A}$ denote matrices and lower-case letters $\boldsymbol{a}$ for vectors. The letter $V$ is reserved for the number of views. The subscript $X_i$ denotes the index of views, $[V]=\{1,\cdots,V\}$ and $X^V=(X_1,\cdots,X_V)$. Additionally, we reserve $S$ for elements of the bipartite index set defined as $\Pi_V=\{s||s|\leq V/2,X^s\cap X^{\bar{s}}=\varnothing,\forall s\subset[V]\}$. For convenience, we will use $X_S$ instead of $X^S$ when no confusion is made. The vector form of discrete probability mass is denoted as:
\begin{IEEEeqnarray}{rCl}
    \boldsymbol{p}_{z|x}=\begin{bmatrix}
        p(z_0|x_0)&\cdots & p(z_0|x_K)&\cdots & p(z_M|x_K)
    \end{bmatrix}^T,\nonumber
\end{IEEEeqnarray}
where $|\mathcal{Z}|=M,|\mathcal{X}|=K$. Similarly, $p_{z|x^V}$ is formed by concatenating elements enumerating the conditioned variables in order.
\section{Common Information for Multiview Learning}
We start with the case where the joint distribution $p(x^V)$ is known but with a hidden common random variable $Y$ where $p(x^V)=\sum_{y\in\mathcal{Y}}p(y,x^V)$ is arbitrarily correlated with $X^V$. Then from the extended Wyner common information formulation in~\cite{huang2024efficient}, we relax the conditional independence constraint in~\cite{wyner1975common} by tolerating a non-zero conditional mutual information threshold instead~\cite{relaxWynerInfo22}:
\begin{IEEEeqnarray}{rCl}
    \underset{P_{z|x^V}\in\Omega_{z|X^V}}{\min}\,&& I(Z;X^V),\IEEEyesnumber\label{eq:mbib_all}\IEEEeqnarraynumspace\IEEEyessubnumber\label{eq:mbib_jointmi}\\
    \text{subject to:}\,&& I(X_S;X_{S^c}|Z)\leq \epsilon_S,\forall S\in\Pi_V.\IEEEeqnarraynumspace\IEEEyessubnumber\label{eq:mbib_relax_cmi}
\end{IEEEeqnarray}
Note that when $\epsilon_S=0,\forall S\in\Pi_V$, we resume $X_i\rightarrow Z\rightarrow X_j,\forall i\neq j\in[V]$.


In~\cite{huang2024efficient}, we solve the unconstrained Lagrangian of~\eqref{eq:mbib_all} with difference-of-convex algorithm, whose main update equation is given by:
\begin{IEEEeqnarray}{rCl}
    P^{(k+1)}_{z|x^V}&=&\frac{P^{(k)}_z}{M^{(k)}}\exp\left\{\sum_{S\in\Pi_V}\kappa_S\sum_{l\in\{S,S^c\}}\log{{P^{(k)}_{x_l|z}}}\right\},\IEEEeqnarraynumspace\IEEEyesnumber\label{eq:wynerdca_eq}\label{eq:wynerdca_pzcx1x2}
\end{IEEEeqnarray}
where $M^{(k)}$ is a normalization constant so that $P^{(k+1)}_{z|x^V}:=p^{(k+1)}(z|x^V)$ is a valid probability mass; $0\leq\ \kappa_S<1,\forall S\in\Pi_V$ and $\sum_{S}\kappa_S<1$. Notably, given a $x^V\in\mathcal{X}^V$, $P(x_l)$ is a constant and does not affect $P_{z|x^V}$.
In~\cite{huang2024efficient}, we apply the above solver to multiview clustering problems for empirical settings, where there is a dataset of $N$ sample pairs of $V$ views, i.e., $D:=\{x_n^V\}_{n=1}^N$. This is achieved through parameterizing the log-likelihoods $P_{z|x_S},\forall S\subset[V]$ as categorical distributions $Cat(q_S(x_S))$ with uniform prior $P_z=1/|\mathcal{Z}|$. Then by Bayes rule, $P_{x_S|z}\propto P_{z|x_S}/P_z$, so the common random variable is a categorical distribution~\eqref{eq:wynerdca_pzcx1x2} with equivalent class probability given by:
\begin{IEEEeqnarray}{rCl}
    q_{eq}(z):=\frac{\prod_{S\in\Pi_V}[q_S(z)q_{S^c}(z)]^{\kappa_S}}{\sum_{z'\in\mathcal{Z}}\prod_{S'\in\Pi_V}[q_{S'}(z')q_{{S'}^c}(z')]^{\kappa_{S'}}}\IEEEeqnarraynumspace\IEEEyesnumber\label{eq:cat_model}.
\end{IEEEeqnarray}
In this case, the common randomness is characterized by the categorical distribution with class probability as shown in~\eqref{eq:cat_model}. However, the method only considers complete view samples and does not apply to incomplete multiview clustering problems.

\section{Generalization to Incomplete Multiview Clustering}
We generalize the proposed common information framework to the practical setting where an incomplete view dataset is given for clustering. Specifically, the missing values could be a different subset of view indices (excluding the case where all views are missing) from one data sample to another. We address the more challenging task by focusing on the available sources of information and directly clustering them. Interestingly, our approach can jointly estimate the cluster target (common randomness) and the missing values in contrast to both imputation and separation methods.

Without loss of generality, consider a sample with incomplete views $x_A$ (unknown values of $x_{\bar{A}}$). We focus on the available incomplete view to estimate the common random variable $z\sim P(z|x_A)$, which corresponds to a cluster distribution, and we partition $x_A$ into a bipartite $x_A=(x_{A_1},x_{A_2}),\forall A_1\in \Pi_A$.

Observe that the components of the exponent of the update equation consist of log-likelihoods $\log P_{z|x_S}, S\subset [V]$. When given incomplete view samples, we can always find the log-likelihood corresponding to $\log{P_{z|x_A}}$, that is, the incomplete views are sufficient to evaluate the log-likelihoods. We achieve this by substituting the likelihoods of the missing value views with the prior probabilities:$P_{z|x_{\bar{A}}}=P_z$, which contribute equally across different cluster distributions in evaluating the log-likelihoods of~\eqref{eq:wynerdca_pzcx1x2}. Mathematically, the update equations are rewritten as:
\begin{IEEEeqnarray}{rCl}
    P^{(k+1)}_{z|x_A}&=&\frac{P^{(k)}_z}{M^{(k)}}\exp\{\kappa_A\log{\frac{P^{(k)}_{z|x_A}}{P^{(k)}_z}}\nonumber\\
    &&+\sum_{a\subset A}\kappa_a[\log{{P^{(k)}_{x_a|z}}}+\log{{P^{(k)}_{x_{\bar{a}}|z}}}]\},\IEEEeqnarraynumspace\IEEEyesnumber\label{eq:incompete_clustering}
\end{IEEEeqnarray}
where $M^{(k)}$ is the step $k$ normalization function and we denote $\bar{a}:=A\backslash \{a\}$. 
In \eqref{eq:incompete_clustering}, the first exponent corresponds to the available-missing bipartite $(A,\bar{A})$, and the second term corresponds to all the bipartite of $A$. Here, we have $\kappa_a=\sum_{S\in\Pi_V}\kappa_{S}\boldsymbol{1}\{S=\{a,\nu\},\exists \nu\in\bar{A}\},a\in A$ with $\boldsymbol{1}\{\cdot\}$ being the indicator function.
It is worth noting that in this case we always have:$P^{(k+1)}_{z|x^V}=P^{(k+1)}_{z|x_A}$. Therefore, the evaluation of the common source is independent of the missing views. In other words, the Markov chain $X_{\bar{A}}\rightarrow X_A\rightarrow Z$ holds, and hence no missing value imputation is required as compared to previous works~\cite{lin2021completer}.

From~\eqref{eq:incompete_clustering}, for each complete/incomplete view sample $n\in[N]$, the equivalent class probability $\forall z\in\mathcal{Z}$, generalizing~\eqref{eq:cat_model} is given by:
\begin{IEEEeqnarray}{rCl}
    q_{eq;n}(z):=\frac{q^{\kappa_A}_{A_n;n}(z)\prod_{a\subset A_n}[q_{a;n}(z)q_{\bar{a}(z);n}]^{\kappa_a}}{\sum_{z'\in\mathcal{Z}}q^{\kappa_A}_{A_n;n}(z')\prod_{a'\subset A_n}[q_{a';n}(z')q_{\bar{a'};n}(z')]^{\kappa_{a'}}},\IEEEeqnarraynumspace\IEEEyesnumber\label{eq:q_incomplete_subset}
\end{IEEEeqnarray}
where $A_n\subset\Pi_V$ is the available index set for the $n$-th sample, and we restrict $P_{z;n}=1/|\mathcal{Z}|,\forall n\in[N]$. The exponents in~\eqref{eq:q_incomplete_subset} have an exponential complexity growth rate (with respect to the number of views) of $\mathcal{O}(2^V)$. One can further impose conditional independence on the feasible solution set:$P_{z|x_A}\propto\prod_{a\in A}P_{z|x_a}$, which reduces the growth rate to linear $\mathcal{O}(V)$ and the class probability in this case is given by:
\begin{IEEEeqnarray}{rCl}
    q^*_{eq;n}(z):=\frac{\prod_{i\in[|A|]}q^{\kappa_{A^*}}_{i;n}(z)}{\sum_{z'\in\mathcal{Z}}\prod_{j\in[|A|]}q^{\kappa_{A^*}}_{j;n}(z')},\IEEEeqnarraynumspace\IEEEyesnumber\label{eq:q_v_cond}
\end{IEEEeqnarray}
where $\kappa_{A^*}:=\kappa_A+\sum_{a\subset{A}}\kappa_a< 1$.

\begin{figure*}[!t]
    \centerline{
        \subfloat[Complete Views]{
            \includegraphics[width=0.25\textwidth]{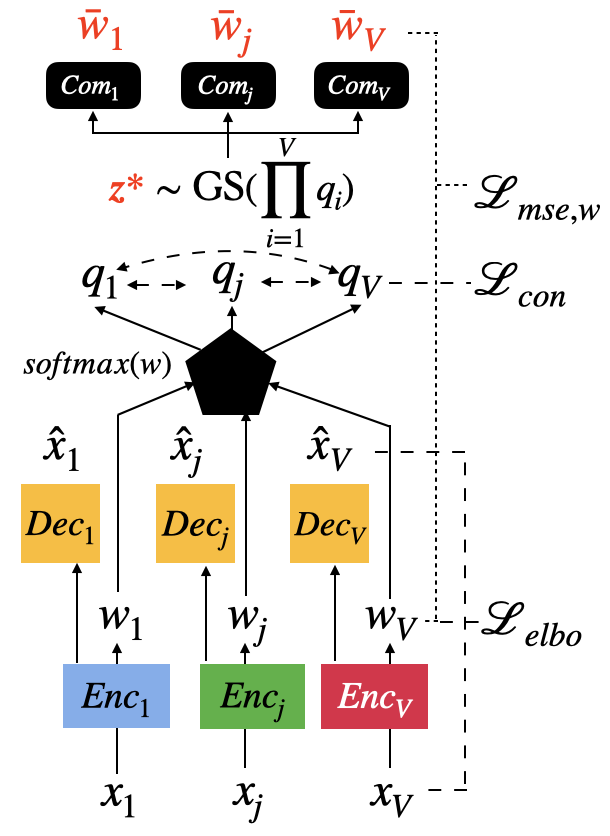}
            \label{subfig:impl_complete}
        }
        \hfil
        \subfloat[No Pair]{
            \includegraphics[width=0.25\textwidth]{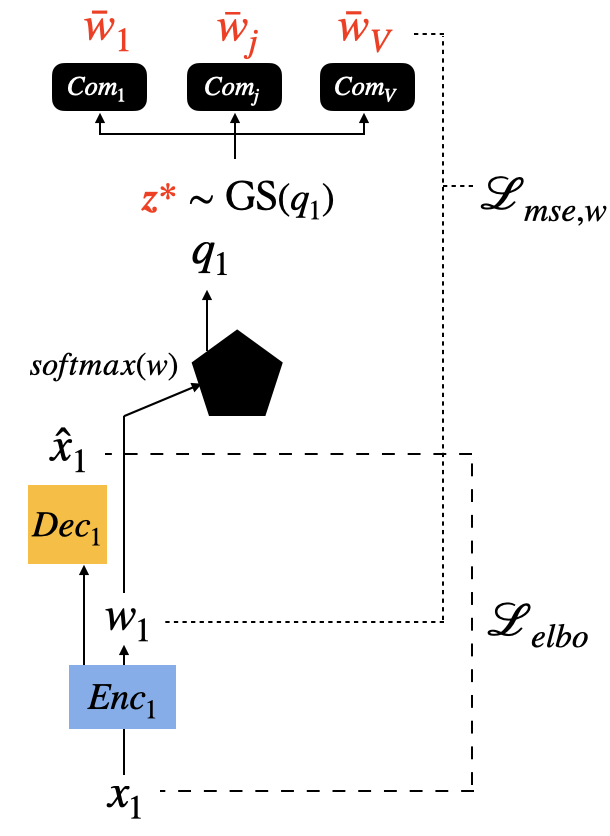}
            \label{subfig:impl_nopair}
        }
        \hfil
        \subfloat[Partially Paired]{
            \includegraphics[width=0.25\textwidth]{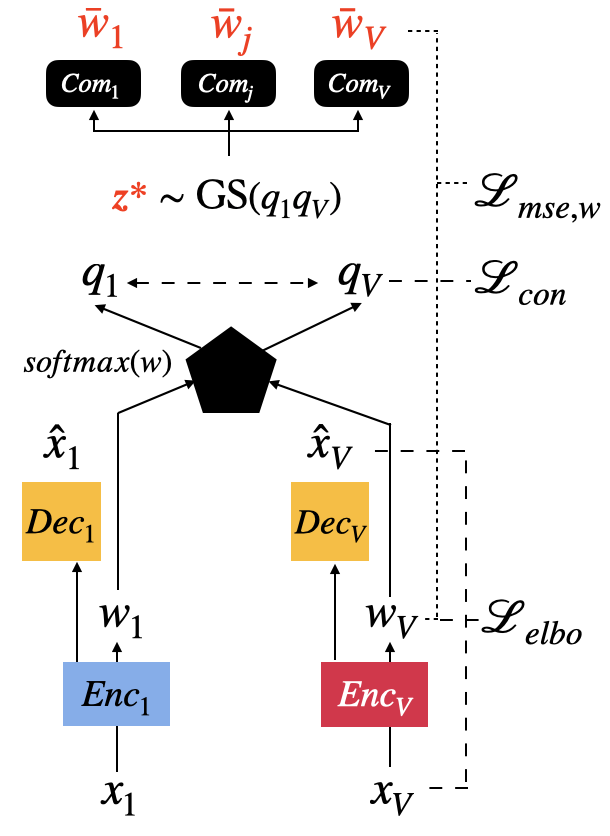}
            \label{subfig:impl_partial}
        }
        }
        \caption{The WyIMVC solver for all sample cases in incomplete multiview clustering. The red letters denote the output of the solvers. The dotted lines denote the objective functions for optimization (See Implementation). $GS(\cdot)$ denotes the adopted Gumbel-softmax reparameterization trick~\eqref{eq:gumbel_softmax}.}
        \label{fig:impl_flow}
\end{figure*}
\section{Implementation}
The objective function of the proposed WyIMVC consists of three main parts: 1) dimension reduction with variational autoencoders, 2) common randomness sampling, and 3) pairwise label matching. For scalability, we implement the conditional independent variant~\eqref{eq:q_v_cond}, but we note that the approach applies to general settings~\eqref{eq:q_incomplete_subset}.

\paragraph{Dimension reduction}
First, for each view observation $X_v$, we compute a low-dimensional $W_v=f_{v;\theta}(X_v)+\varepsilon, \hat{X}_v=g_v(W_v)$ where $f_v,g_v$ are implemented with neural networks with parameters $\theta$, and $\varepsilon\sim\mathcal{N}(0,\boldsymbol{I})$. We minimize the evident lower bound~\cite{kingma2013auto}, given by:$\mathcal{L}^{v}_{elbo}=\mathcal{L}^v_{mse}+\mathcal{L}^v_{KL}$.
The reconstruction part is defined as:
\begin{IEEEeqnarray}{rCl}
    \mathcal{L}^{(v)}_{mse}=\frac{1}{N_v}\sum_{n=1}^{N}m^{(n)}_v\lVert \boldsymbol{x}^{(n)}_v-g_{v;\theta}(\boldsymbol{w}_v))\rVert_2^2,
\end{IEEEeqnarray}
where $m^{(n)}_v\in\{0,1\}$ is the indicator of missing values (case $0$), and $1$ otherwise; $N_v=\sum_{n=1}^Nm_v^{(n)}$. As for the Kullback-Leibler divergence, since here we parameterize the neural networks as $\mathcal{N}(f_v(\boldsymbol{x}_v),\boldsymbol{I})$, with reference density function $\mathcal{N}(0,\boldsymbol{I})$, the closed-form expression is given by:
\begin{IEEEeqnarray}{rCl}
    \mathcal{L}_{KL}^{(v)}=\frac{1}{N_v}\sum_{n=1}^N m_v^{(n)}\lVert f_{v;\theta}(\boldsymbol{x}^{(n)}_v)\rVert_2^2,
\end{IEEEeqnarray}
where we omit the constants independent of optimization parameters. Our implementation differs from~\cite{xu2024deepvae}: in our method, the outputs of the variational encoders are not treated as common features, whereas in~\cite{xu2024deepvae} they are used to form the jointly Gaussian means and variances for common information sampling.

\paragraph{Common randomness sampling}


The common random variable $Z$ in our solver coincides with the cluster targets (cluster distributions), so the Gaussian reparameterization trick does not apply here~\cite{kingma2013auto}. To avoid the intractability, we adopt the Gumbel-softmax reparameterization trick~\cite{jang2016categorical}, which allows for a well-defined gradient for optimization. The definition of a $d$ dimensional Gumbel-softmax distribution is given by:
\begin{IEEEeqnarray}{rCl}
    g_{\pi,\tau}(\boldsymbol{z}):=\Gamma(d)\tau^{d-1}(\sum_{i=1}^d\pi_i/{z_i^\tau})^{-d}\prod_{j=1}^d(\pi_i/{z_i^{\tau+1}}),
\end{IEEEeqnarray}
where $\Gamma$ is the Gamma function, $\boldsymbol{z}^\tau$ belongs to the probability simplex, $\pi_i,\forall i\in[d]$ denotes the category distributions, and $\tau>0$. 
The Gumbel-softmax distribution can be sampled as follows:
\begin{IEEEeqnarray}{rCl}
    z^{\tau}_i=\frac{\exp\{(\log{\pi_i} + s_i)/\tau\}}{\sum_{d}\exp\{(\log{\pi_j}+s_j)/\tau\}},\IEEEeqnarraynumspace\label{eq:gumbel_softmax}
\end{IEEEeqnarray}
where $s_i\sim\text{Gumbel(0,1)}$ defined as $\text{Gumbel}(\mu,\beta):=e^{-e^{-(s-\mu)/\beta}}$. In sampling $s_i$, one performs inverse transform sampling with a uniformly distributed source $\mu_i\sim\mathcal{U}(0,1),\forall i\in[d]$, followed by: $s_i=-\log{(-\log{\mu_i})}$.
We simply substitute $\pi_i$ in~\eqref{eq:gumbel_softmax} with~\eqref{eq:q_v_cond} for Gumbel-softmax sampling of the $n$-th sample.

Given the realizations of the common random variable, we can infer the missing values with an extra set of neural networks: $\hat{\boldsymbol{w}}^{(n)}_v=h_{v;\theta}(\boldsymbol{z}^{(n)}),\forall v\in[V]$, and we again impose MSE loss for optimization:
\begin{IEEEeqnarray}{rCl}
    \mathcal{L}^{(v)}_{mse,w}=\frac{1}{N_v}\sum_{n=1}^{N_v}m_v^{(n)}\lVert \boldsymbol{w}_v^{(n)}-h_{v;\theta}(\boldsymbol{z}^{(n)})\rVert^2_2.
\end{IEEEeqnarray}

\paragraph{Label Matching with Complete Samples}
Next, we align the view-specific class probability $\boldsymbol{q}_v^{(n)},\forall v\in[V]$ using the contrastive learning method~\cite{chen2020simple,do2021clustering,xu2022multi}. This is performed on each pair of $\{\boldsymbol{q}^{(n)}_v,\boldsymbol{q}_{w}^{(n)}\}_{n\in[N]},\forall v\neq w\in[V]$. Then we focus on complete pairs, i.e.,$U_{v,w}:=\{n|m_v^{n}= m_w^{n}=1,\forall n\in[N]\}$, and concatenate the sample-wise predictions as a matrix $\boldsymbol{C}_v:=\begin{bmatrix}q_v^{u_1}&\cdots &q_v^{u_{|U|}}\end{bmatrix}^T$, which are used to compute the positive examples:
\begin{IEEEeqnarray}{rCl}
    {\Lambda}_+^{(i,j)}=\{\lambda|\lambda\in \text{diag}[\cos(\boldsymbol{C}_i^{u},\boldsymbol{C}_j^u)],\forall i,j\in\{v,w\},u\in U_{v,w}\},\nonumber
\end{IEEEeqnarray}
where $\cos(\boldsymbol{A},\boldsymbol{B})_{i,j}=\frac{\boldsymbol{a}^T_i\boldsymbol{b}_j}{\lVert \boldsymbol{a}_i\rVert_2\lVert \boldsymbol{b}_j\rVert_2}$ is the element-wise cosine similarity function. Similarly, the negative examples are defined as:
\begin{IEEEeqnarray}{rCl}
    \Lambda_-^{(i,j)}=\{\lambda|\lambda\notin \text{diag}(\boldsymbol{C}_i^u[\boldsymbol{C}_j^u]^T),\forall i,j\in\{v,w\},u\in U_{v,w}\}.\nonumber
\end{IEEEeqnarray}
With the definitions, we apply the multiview contrastive learning loss on pairwise complete samples:
\begin{IEEEeqnarray}{rCl}
    \mathcal{L}_{con}^{(i,j)}=-\sum_{\lambda_+\in\Lambda_+^{(i,j)}}\log{\frac{\lambda_+}{\lambda_+ + \sum_{\lambda_-\in\Lambda_{-}^{(i,j)}}{\lambda_-}}},
\end{IEEEeqnarray}
\paragraph{Overall objective function}
Summarising the above, the overall objective function is given by:
\begin{IEEEeqnarray}{rCl}
\bar{\mathcal{L}}=\sum_{v=1}^V\mathcal{L}^v_{elbo}+\mathcal{L}_{mse,w}^{(v)}+\sum_{i=1}^V\sum_{j=1}^V\mathcal{L}_{con}^{(i,j)}.
\end{IEEEeqnarray}
 
The computation of the objective function depends on the available views; all possible scenarios are summarized in Fig.~\ref{fig:impl_flow}. Finally, the differences from the compared solvers are in order:
\begin{itemize}
    \item The solver~\cite{huang2020multimodal} applies to the complete view settings, i.e., $m_v^{(n)}=1,\forall v\in[V], n\in[N]$. Additionally, we adopt variational autoencoders for dimension reduction, whereas deterministic autoencoders are adopted in~\cite{huang2024efficient}.
    \item In contrast to all compared methods, we adopt the Gumbel-softmax reparameterization trick with the equivalent class distribution following the DCA insights. Therefore, our solver obtains the realization of the common random variable $\{z_n\}_{n=1}^N$ for each sample pair. Additionally, the realizations $\{z_n\}_{n=1}^N$ allow missing value imputation, whereas the compared methods output cluster predictions only.
\end{itemize}
\begin{figure}[!t]
    \centering
        \subfloat[Handwritten Dataset]{
            \includegraphics[width=3in]{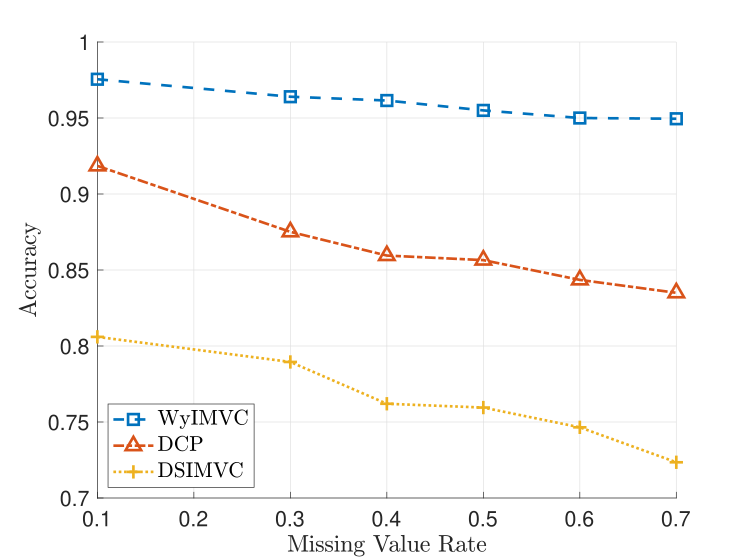}
            \label{subfig:acc_hand}
        }
        \\
        \subfloat[Fashion Dataset]{
            \includegraphics[width=3in]{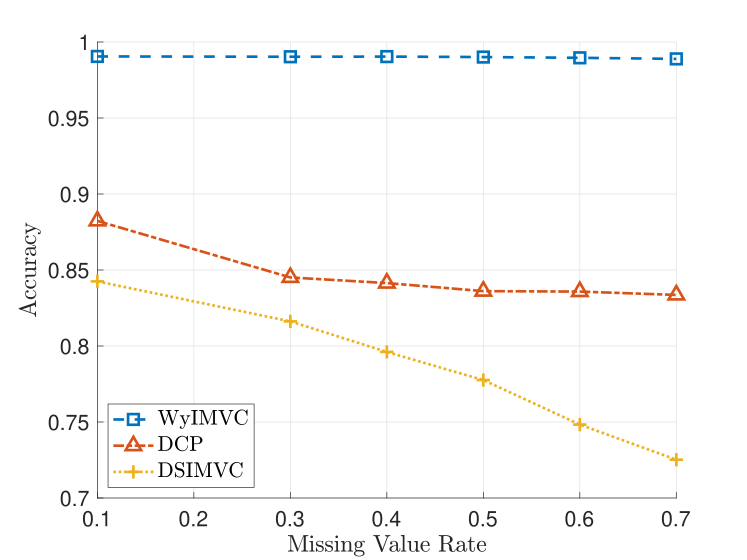}
            \label{subfig:acc_fashion}
        }
        \\
        \subfloat[MSRCv1 Dataset]{
            \includegraphics[width=3in]{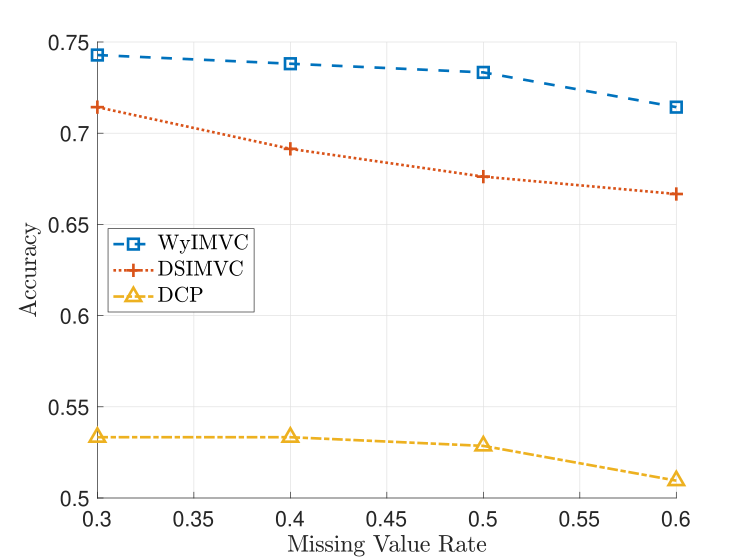}
            \label{subfig:acc_msrcv1}
        }
        \caption{Clustering accuracy versus proportion of samples with missing views for the evaluated datasets.}
        \label{fig:clustering_acc}
\end{figure}

\section{Evaluation}
\subsection{Datasets}
We use three non-trivial datasets to demonstrate the advantages of our solver:
\begin{itemize}
\item Handwritten~\cite{uci_repo}: A digit recognition dataset consisting of $6$ features. This dataset contains $2,000$ samples of handwritten digits from $0-9$.
\item Fashion~\cite{xiao2017fashionmnistnovelimagedataset}: Three different styles of a product are paired together to create $10,000$ samples, $3$ views, and $10$ clusters image dataset.
\item MSRC-v1~\cite{winn2005locus}:  For scene recognition task, a multi-view dataset with $210$ samples categorized into $7$ classes is created. Each scene is represented using $5$ views.
\end{itemize}
The details of the datasets are listed in Table~\ref{tab:datasets}. These datasets are complete multiview datasets, and we manually modify them to incomplete ones. First, we uniformly select a portion of samples (scale $(0.0,1.0]$ with $1.0$ means all samples). Then, we randomly choose a case uniformly over all subsets of views (excluding the all-missing-value case), and mask the values of the views in the selected subset. We call the parameter of the configuration ``missing value rate'', which we vary in the range $(0.1,0.7]$, e.g., $0.1$ means $10\%$ of the data samples have missing values in a subset of all views.

\begin{table}[!b]
\caption{Datasets Statistics}
\label{tab:datasets}
\centering
\begin{tabular}{ccccc}
\hline
\textbf{Datasets} & \textbf{Samples} & \textbf{Views} & \textbf{Clusters} & 
\\ \hline\hline
Handwritten      & 2,000     & 6      & 10
\\ \hline
Fashion          & 10,000    & 3      & 10
\\ \hline
MSRC-v1          & 210       & 5      & 7
\\ \hline
\end{tabular}
\end{table}

\subsection{Compared methods}
We compare the proposed WyIMVC to two state-of-the-art incomplete multiview clustering algorithms:
\begin{enumerate}[label=\roman*)]
\item Dual Contrastive Prediction for Incomplete Multiview Representation Learning (DCP-CV)~\cite{lin2022dual,lin2021completer}: The solver uses contrastive learning to maximize the mutual information between the different views and then recover the missing views by minimizing the conditional entropy.
\item Deep Safe Incomplete Multi-view Clustering (DSIMVC)~\cite{tang2022deep}: The authors proposed a bi-level optimization framework where missing views are reconstructed and selected for training by leveraging learned semantic neighbors to mitigate the risk of clustering performance degradation caused by semantically inconsistent imputed views.
\end{enumerate}
We use their latest available source codes with minimal modifications for a fair comparison (same network architecture for autoencoders, mini-batch sizes for stochastic gradient descent as in~\cite{chen2023deep})\footnote{DCP-CV baseline is available at https://github.com/XLearning-SCU/2022-TPAMI-DCP; DSIMVC baseline is available at https://github.com/Gasteinh/DSIMVC}. The compared methods all attempt to use information-theoretic objective functions for optimization and use the shared information for incomplete multiview clustering. Contrary to these approaches, our method directly clusters the available views, which essentially estimates the common randomness, with the realization obtained by Gumbel-softmax sampling.

In each experiment, the parameters of a solver are initialized with a random seed. Then, they are optimized for $400$ epochs using the proposed objective function of each method. Afterward, we record the cluster predictions. We repeat the procedures for $5$ different random seeds. The clustering accuracy is computed offline and is done by a standard label matching processing with the ground truth cluster labels~\cite{labelmatching}. 
\subsection{Results}
In Fig.~\ref{fig:clustering_acc}, we report the clustering accuracy of the compared methods on the three evaluated datasets. In each plot, we compare the obtained accuracy of each solver, varying the missing value rate of a dataset. In Fig.~\ref{subfig:acc_hand}, our method outperforms the compared solvers over the evaluated range of missing value rate. Notably, the degradation against increased missing values is less severe as compared to both baselines. This demonstrates the strength of the proposed solver. In Fig.~\ref{subfig:acc_fashion} and Fig.~\ref{subfig:acc_msrcv1}, similar trends are observed.

\section{Conclusion}
To conclude, by generalizing the recent trends in common information-based multiview learning, we extend the Wyner common information paradigm to incomplete multiview clustering problems. We leverage the DC structure of the formulated problem for efficient optimization. Our solver, implemented with neural networks, can directly estimate the common randomness by an efficient Gumbel-softmax sampling strategy. Moreover, our empirical evaluation demonstrates improved clustering performance against the compared methods. 

For future work, one issue that is not addressed, due to the scope of this work, is the selection of ``mixture parameter'', e.g., $\{\kappa\}_S$ in~\eqref{eq:q_v_cond}. It is closely related to Bayes Fisher information~\cite{8502792}, and the generalized escort distribution in information divergence and statistical physics~\cite{10547443}. A rigorous study of this connection will be left for future exploration.

\bibliographystyle{IEEEtran}
\bibliography{references}

\appendices

\end{document}